\documentstyle[11pt,newpasp,twoside,epsf]{article}
\def\edcomment#1{\iffalse\marginpar{\raggedright\sl#1\/}\else\relax\fi}
\reversemarginpar

\begin{document}

\title{Mapping the Remote Milky Way Halo using BHB stars at 70 $<$ r $<$ 130 kpc}
 \author{L. Clewley}
\affil{Department of Physics, Denys Wilkinson Bldg., University of Oxford, Keble Road, Oxford OX1 3RH}
\author{S. J. Warren}
\affil{Blackett Laboratory, Imperial College of Science Technology and Medicine, Prince Consort Rd, London SW7 2BW}
\author{P. Hewett, M. Wilkinson, W. Evans}
\affil{Institute of Astronomy, Madingley Road, Cambridge CB3 0HA}

\begin{abstract}
We increase the number of remote halo tracers by using blue
horizontal branch (BHB) stars out to Galactocentric distances of 130
kpc. We use SDSS EDR photometry and the VLT to detect 16 BHB stars at
Galactocentric distances 70 $< $r $<$ 130, and to measure their radial
velocities. We find the mass of the Milky Way is $M =1.7^{+3.0}_{-0.6}
\times 10^{12} M_{\odot}$. When completed this survey will: (i)
substantially reduce the errors in the total mass and extent of
the Milky Way halo, (ii) map the velocity space in a hitherto
unexplored region of the halo. 
\end{abstract}

\section{Introduction}
The total masses and sizes of all galaxies are poorly determined
quantities \---\ because we do not have suitable dynamical tracers at
sufficiently large radii. The accurate measurement of the mass profile
provides important clues to the nature of the dark matter. The Milky
Way is the prime target for an accurate measurement. Wilkinson \&
Evans (1999, hereafter WE99) calculate the total mass of the Milky Way
to be $M_{tot}=1.9^{+3.6}_{-1.7} \times 10^{12} M_{\odot}$, using the
full set of 27 known satellite galaxies and globular clusters
(hereafter satellites) at Galactocentric radii $r>$ 20 kpc (six
possess measured proper motions). The large errors are primarily a
consequence of the small number of satellites. The sample must be
nearly complete, so a new population of distant halo objects must be
found in order to increase the number of dynamical tracers. This
motivates a new survey for remote halo tracers at large Galactocentric
distances.

\section{Isolating BHB stars at $70 < r < 130$ kpc}
We have solved the problem of how to find large numbers of BHB stars
in the remote halo, and have begun a VLT programme to measure the
kinematics of the population. The procedure involves three
steps. First, we use u*g*r* photometry from the SDSS database to
detect A-type stars at high-Galactic latitude. These stars are either
BHB stars ($M_B$ = 0.7, luminous standard candles) or blue stragglers
which are $\sim$ 2 mag. less luminous. Second, we
reduce the number of blue stragglers and increase the probability of
good classification by limiting the colour ranges to 1.4 $< u*-g* <$
1.1 and -0.04 $> g*-r* >$ -0.2. Finally, using spectroscopy, described
in Clewley et al. (2002), we are able to efficiently separate the two
populations. We consistently find that about half of our target A-type
stars are BHB stars. Figure 1 (middle) shows the increase of distant
halo tracers at Galactocentric distances of 20 $< r <$ 130 kpc from 23
(left) to 90.
\begin{figure}
\plotone{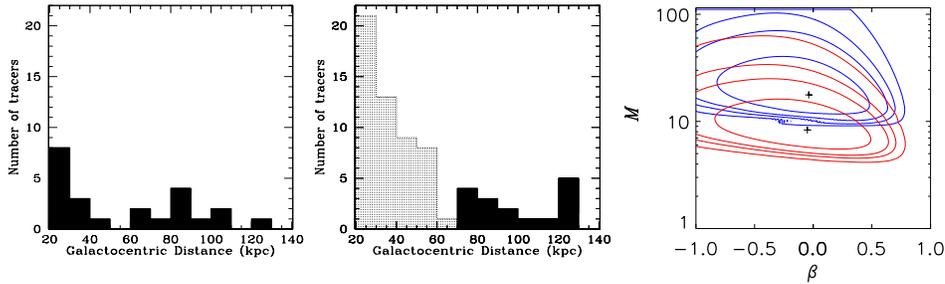}
\caption{The {\em left plot} is a histogram of the 23 satellites at
  Galactocentric distances of 20 $< r <$ 130 kpc. The {\em middle
  plot} shows 51 BHB stars at distances 20 $< r <$ 70 kpc made from
  our previous study (Clewley, 2002) and 16 BHB stars at 70 $< r <$ 130
  kpc made using SDSS and the VLT. The {\em right plot} shows the resultant
  likelihood contours for the mass, M (in $10^{11}M_{\odot}$) and the
  velocity anisotropy, $\beta$ for the combined data shown in both histograms
  using the WE99 model. Contours are enclosed probabilities of 68\%,
  90\%, 95\% and 99\% including Leo I (upper curves) and
  excluding Leo I (lower curves).}
\end{figure}
\section{The mass of the Milky Way}
Preliminary results from the VLT are extremely encouraging. We have
successfully isolated 16 BHB stars in the Galactocentric distance
range of 70 to 130 kpc from a target sample of 31 candidates. With
this small programme we have nearly doubled the number of tracers
known at $r>70$kpc.  Combined with the 23 satellites and 51 BHB stars
at distances 20 to 60 kpc (from previous work at the AAT and WHT) we
find the mass of the Milky Way is $M =1.7^{+3.0}_{-0.6} \times 10^{12}
M_{\odot}$.

\section{References}
Clewley, L. Warren, S. J., Hewett P. C., Norris John E., Peterson,
R. C., Evans, N. W., 2002, MNRAS, 337,87 \\
Clewley L., 2002, PhD thesis, University of London \\
Wilkinson M., Evans N., 1999, MNRAS, 310, 645 \\
\end{document}